\documentclass[aps,prl,twocolumn,amsmath,amssymb,showpacs]{revtex4-1}
\usepackage{graphicx}
\usepackage{dcolumn}
\usepackage{bm}

\usepackage{latexsym}
\usepackage{amsthm}
\usepackage{braket}
\allowdisplaybreaks

\usepackage[top =1in,bottom = 1in, left = 1in, right=1in]{geometry}

 \newcommand{\arXiv}[1]{\href{http://www.arXiv.org/abs/#1}{arXiv:#1}}
\usepackage[colorlinks=true, linkcolor=blue, bookmarks=true]{hyperref}

\makeatletter
\renewcommand\section{\@startsection {section}{1}{\z@}%
                  {-3.5ex \@plus -1ex \@minus -.2ex}
                  {2.3ex \@plus.2ex}%
                  {\normalfont\large\bfseries}}
\renewcommand\subsection{\@startsection{subsection}{2}{\z@}%
                   {-3.25ex\@plus -1ex \@minus -.2ex}%
                   {1.5ex \@plus .2ex}%
                   {\normalfont\bfseries}}
\makeatother

\newcommand{\beq}{\begin{equation}}
\newcommand{\eeq}{\end{equation}}
\newcommand{\ber}{\begin{array}}
\newcommand{\eer}{\end{array}}
\newcommand{\del}{\partial}

\newcommand{\dsty}{\displaystyle}
\newcommand{\te}{\theta}

\newcommand{\de}{\delta}

\newcommand{\om}{\omega}

\newcommand{\ena}{\end{eqnarray}}
\newcommand{\beqa}{\begin{eqnarray}}
\newcommand{\eeqa}{\end{eqnarray}}
\newcommand{\bea}{\begin{eqnarray}}
\newcommand{\eea}{\end{eqnarray}}

\newcommand{\al}{\alpha}
\newcommand{\ald}{\alpha^\dagger}

\theoremstyle{remark}

\begin{document}

\title{\Large Spectroscopy instead of scattering:\\ particle experimentation in AdS spacetime}
\author {\large Oleg Evnin\vspace{3mm}}

\affiliation{
Department~of~Physics,~Faculty~of~Science,~Chulalongkorn~University,~Bangkok,~Thailand\\
\& Theoretische Natuurkunde (VUB) \& International Solvay Institutes, Brussels, Belgium}

\begin{abstract}
Particle experiments are difficult at weak coupling because interactions are rare and a huge number of collision attempts are needed to attain significant precision. One often hears that `one Higgs boson is produced in a billion of collisions at LHC.' In this essay, we fantasize about possible advantages afforded in this regard by performing experiments in anti-de Sitter (AdS) spacetime instead of a usual collider in a nearly-flat spacetime. Being a perfectly resonant cavity, the AdS spacetime enhances all nonlinear interactions, which therefore produce effects of order one no matter how small the couplings are, provided that one waits long enough. These effects are encoded in spectroscopic data,  namely, the fine splittings of the energy levels which would have been highly degenerate in AdS if no interactions were present. Over long times, such energy shifts let different components of wavefunctions drift completely out-of-phase, producing large effects for arbitrarily small interactions.

\end{abstract}

\maketitle

\noindent{\bf Introduction:} Experimental particle physics is difficult. One constructs devices of geographic dimensions, employs thousands of people, writes  volumes of software, carefully prepares the beams, collides them -- and most of the time, nothing of interest happens! A slogan is often heard that `one Higgs boson is produced in a billion of collisions at LHC' -- some more precise analysis can be found in \cite{higgs}. In this essay, we shall pursue a hypothetical scenario where a particle experimentalist would have access to anti-de Sitter (AdS) spacetime. While there is admittedly no foreseeable way to implement this setup, it offers an interesting cure to contemplate for low interaction rates. Being a perfectly resonant cavity, AdS boosts arbitrarily small interactions to effects of order 1 if one waits long enough. It is instructive to imagine how such arrangements would  function.

AdS is a maximally symmetric spacetime that has exactly as many symmetries as the Minkowski spacetime (and as such, it is special even from a purely mathematical standpoint). It emerges in Einsteinian gravity as the vacuum solution with a negative cosmological constant. It plays a special role in the much-studied AdS/CFT correspondence -- see, e.g., \cite{AdS}. The metric of the 4-dimensional AdS$_4$ spacetime reads
\beq
ds^2=\frac{L^2}{\cos^2x}\left[-dt^2+dx^2+\sin^2x\left(d\te^2+\sin^2\te\, d\varphi^2\right)\right].
\label{adsmetric}
\eeq
The AdS radius $L$ sets the spacetime curvature and is related to the cosmological constant as
\beq
\Lambda=-\frac3{L^2},
\eeq 
the radial coordinate $x$ runs between 0 and $\pi/2$, and $\te$ and $\varphi$ are the usual 2-sphere angles. In the treatment below, we shall choose units such that $L=1$. Even though translation symmetries are not manifest in our coordinates, all points and directions are equivalent, as they are in Minkowski space, and there are also analogs of boosts. The spatial distance between $x=0$ and $x=\pi/2$ is infinite, but the finite range of this variable creates the correct impression that, dynamically, AdS operates as a cavity confining the fields -- and a very special sort of cavity indeed, as we shall repeatedly see below.

While at the level of symmetries, AdS shows a lot of resemblance to the familiar Minkowski space, as an arena for dynamics, it couldn't be more different. Dynamics in Minkowski is characterized by scattering and dispersal to infinity. If no reaction has happened in the first go, the ingredients disperse and move freely till the end of time. Not so in AdS. The gravitational field of AdS turns all mater waves around and makes them collide repeatedly infinitely many times. Interactions thus have infinitely many attempt to take place. As a signature example of this type of dynamics, all solutions of the wave equation in AdS are exactly periodic in time, with the same period. More generally, there are rigid resonant structures in the eigenmode frequency spectrum in AdS for fields of any spin or mass. For scalar fields, this pattern will be explicitly displayed below.  AdS thus acts as a perfectly resonant cavity that neatly prevents any dispersal.

The resonances are responsible for a dramatic enhancement of interactions. While this will be apparent at a precise level from our subsequent discussion, intuitively, this comes about from the AdS gravitational field turning the dispersing reaction products back and forcing them to collide infinitely many times. For a given small coupling $\lambda$, waiting for times $\sim 1/\lambda$ produces effects of order 1 for arbitrarily small interactions. Could one try to make use of this interaction enhancement in a perfectly resonant cavity? Let's study this question in more detail using the most basic example of a self-interacting quantum scalar field in AdS.

 
\noindent{\bf Quantum fields in AdS:} As the simplest toy model, consider a scalar field with quartic sel-interactions in AdS with the action
\beq\label{Lagr}
S=\int  \left[(\del\phi\cdot\del\phi)+m^2\phi^2+\lambda\phi^4\right] \sqrt{-g}\,\,d^4 x,
\eeq
where the dot-product and the integration measure are constructed with the metric (\ref{adsmetric}). We shall briefly discuss below more complicated field systems, but this simplest toy model serves as a convenient illustration. (Note that it has recently been considered in \cite{BSS} from a different, AdS/CFT-related perspective.)

If we were in Minkowski space instead, the story is well-known and is a basic example from textbooks \cite{PS}. At $\lambda=0$, the solutions are plane waves quantized into free relativistic particles. When lambda is turned on, at linear (tree) level, one computes the matrix elements of the interaction term $\phi^4$ between these particle states to obtain scattering amplitudes, which are then squared and converted into interaction cross-sections.

The story in AdS is rather different. Because of the cavity-like properties of AdS, the linearized solutions are standing waves. Scattering does not exist in cavities, and spectroscopy is the natural language. One may talk about the spectrum of energies in our AdS cavity, and examine how it is affected by the interactions. The Schr\"odinger picture of the elementary quantum mechanics textbooks provides a convenient framework for such discussions.

Let's put our words into formulas. The Hamilatonian corresponding to (\ref{Lagr}) is
\beq\label{H}
\hat{\cal H} = \int  \Big[\cos^2\hspace{-0.7mm}x\hspace{1mm}(\hat{\dot\phi}^2+\nabla\hat\phi\cdot\nabla\hat\phi)+m^2\hat\phi^2+\lambda\hat\phi^4\Big] \,d^3 {\bf x},
\eeq
where the Schr\"odinger picture operators only depend on the spatial coordinates ${\bf x}=(x,\te,\varphi)$, while the dot-product is computed with the (3-sphere) metric $dx^2+\sin^2\hspace{-0.5mm}x\hspace{0.5mm}(d\te^2+\sin^2\te\, d\varphi^2)$ and $d^3 {\bf x}=dx\,d\te\,d\varphi \sin^2x\sin\te/\cos^4x$. We then follow the standard route introducing the creation-annihilation operators $\alpha^\dagger_{nlk}$ and $\alpha_{nlk}$ satisfying
\beq
[\alpha^\dagger_{nlk},\alpha_{n'l'k'}]=-\de_{nn'}\de_{ll'}\de_{kk'},
\eeq 
one pair for each normal mode of $\phi$ in the AdS cavity, labelled by three integers: the usual 3-dimensional angular momentum numbers $l$ and $k=-l,\ldots,l$, with the square of the angular momentum given by $l(l+1)$, and the (nonnegative) radial overtone $n$. One writes
\begin{align}
&\hat\phi=\sum_{nkl}\frac1{\sqrt{\om_{nkl}}}\left(\alpha_{nlk}+\alpha^\dagger_{nlk}\right) e_{nlk}({\bf x}),\label{phiexp}\\
&\hat{\dot\phi}=i\sum_{nkl}\sqrt{\om_{nkl}}\left(\alpha_{nlk}-\alpha^\dagger_{nlk}\right) e_{nlk}({\bf x}).
\end{align}
Here,  $e_{nlk}({\bf x})$ are the AdS mode functions verifying
\begin{align}
&\Bigg[\frac1{\tan^2x}\del_x(\tan^2x\,\del_x)+\frac{\Delta_2}{\sin^2x}-\frac{m^2}{\cos^2x}\Bigg]e_{nlk}\\
&\nonumber\hspace{6cm}=-\om^2_{nlk}e_{nlk},\\
&\hspace{1cm}\Delta_2 e_{nlk}=-l(l+1)e_{nlk},\nonumber
\end{align}
where $\Delta_2$ is the 2-sphere Laplacian.
Explicit expressions (in terms of Jacobi polynomials and spherical harmonics $Y_{lk}$) can be found, say, in \cite{Yang, EN}. The corresponding frequencies are given by 
\beq
\om_{nlk}=\de+2n+l, \qquad \de= \frac3{2}+\sqrt{\frac{9}4+m^2}.\label{freqdef}
\eeq 
Importantly, the difference of any two frequencies is integer. This {\it resonant} property is intimately linked to the symmetries of AdS \cite{EN} and persists in general for fields of arbitrary spins of masses. Note that $m^2$ in (\ref{freqdef}) really means $m^2L^2$, a dimensionless combination of the scalar field mass and the AdS radius $L$, but we have agreed to set $L$ to 1 by a choice of units. 

Substituting the above expansions into (\ref{H}), we obtain (after subtracting the ground state energy)
\beq
\hat{\cal H}=\sum_{nlk} \om_{nlk} \al^\dagger_{nlk}\al_{nlk} + \lambda\,\hat{\cal H}_{int},\label{Hdecmp}
\eeq
where $\hat{\cal H}_{int}$ is simply $\int \hat\phi^4$ expressed through (\ref{phiexp}), namely
\begin{align}
&\nonumber\hat{\cal H}_{int}=\hspace{-5mm}\sum_{\eta_1,\eta_2,\eta_3,\eta_4} \hspace{-5mm}C_{\eta_1\eta_2\eta_3\eta_4}(\al_{\eta_1}+\al^\dagger_{\eta_1})(\al_{\eta_2}+\al^\dagger_{\eta_2})\\
&\hspace{2cm}\times(\al_{\eta_3}+\al^\dagger_{\eta_3})(\al_{\eta_4}+\al^\dagger_{\eta_4}),\label{Hintdef}\\
&\nonumber C_{\eta_1\eta_2\eta_3\eta_4}=\frac{\int e_{\eta_1}({\bf x})\,e_{\eta_2}({\bf x})\,e_{\eta_3}({\bf x})\,e_{\eta_4}({\bf x}) \,d^3 {\bf x}}{\sqrt{\om_{\eta_1}{\om_{\eta_2}}{\dsty\om_{\eta_3}}{\om_{\eta_4}}}},\rule{0mm}{7mm}
\end{align}
where we have introduced the collective index $\eta=(n,l,k)$ so that $\sum_\eta=\sum_{n,l,k}$, $\om_\eta=\om_{nlk}$, $\eta_1=(n_1,l_1,k_1)$, etc.

Now, if $\lambda=0$ one simply has a collection of decoupled oscillators. The eigenstates are given by the Fock basis $|\{n\}\rangle$ which has definite occupation numbers $n_\eta$ for all AdS normal modes:
\beq
\ald_\eta\al_\eta|\{n\}\rangle = n_\eta |\{n\}\rangle.
\eeq
The energies of such states in a non-interacting system are given by
\beq\label{Edef}
E=\sum_\eta \om_\eta n_\eta.
\eeq
Note that the energy levels are highly degenerate. Introduce the total particle number
\beq
N=\sum_\eta n_\eta.
\eeq
Then, 
\beq\label{Epdef}
E'=E-\de N
\eeq
is integer, being made of integers $\om_\eta-\de$ and $n_\eta$. Of course, there are many different ways to generate a given integer value of $E'$ by choosing $n_\eta$. Note that if $\de$ is not a generic integer number, but is rational, extra degeneracies in $E$ occur between different values of $N$ (this does not happen for generic real $\de$ where all degenerate states have the same $N$). However, in those cases there are selection rules satisfied by $C$ in (\ref{Hintdef}) that make the matrix elements of the perturbation Hamiltonian (\ref{Hintdef}) vanish unless it is computed between states with the same $N$. These selection rules have been studied at length in the literature \cite{Yang,EN} for the case of a massless field, which corresponds in AdS$_4$ to $\de=3$ (equal to the number of spatial dimensions). Effectively, we can thus ignore this subtlety in our subsequent analysis of level splitting and assume that the degenerate levels of the unperturbed system are labelled by two integers, $N$ and $E'$. Physically, this means that there will be no particle production due to resonances between the AdS normal modes when we turn on the interactions (interactions only induce mixing between unperturbed states with the same particle numbers). \vspace{3mm}


\noindent{\bf Energy shifts:} When a small nonzero $\lambda$ is turned on in (\ref{Hintdef}), the degenerate energy levels we have just described split.The analysis of this splitting at linear order in $\lambda$ is standard and known from textbooks on quantum mechanics. One must simply compute the matrix elements of $\hat{\cal H}_{int}$ between the states {\it within the same} unperturbed energy level, and diagonalize the resulting finite-sized matrices.

Because we are only interested in the matrix elements of $\hat{\cal H}_{int}$ between states with the same $N$ and $E'$ (and hence the same $E$), the expression (\ref{Hintdef}) can actually be simplified. First, since acting on a given ket-state, we must obtain a state with the same number of particles, only terms with two $\al$'s and two $\ald$'s can contribute. Taking into account the permutation symmetries of $C$, this gives
\beq
\hat{\cal H}_{int}=\hspace{-4mm}\sum_{\eta_1,\eta_2,\eta_3,\eta_4} \hspace{-4mm}C_{\eta_1\eta_2\eta_3\eta_4} \ald_{\eta_1}  \ald_{\eta_2}  \al_{\eta_3}  \al_{\eta_4}.
\eeq
(We shall ignore any overall numerical factors that can be absorbed into a redefinition of $\lambda$.) Second, since acting on a given ket-state, one must obtain a state with the same value of $E'$ (and hence $E$), the total contribution to $E$ removed by the two annihilation operators must be the same as the total contribution added by the two creation operators, i.e., $\om_1+\om_2=\om_3+\om_4$, where $\om_i$ are the frequencies (\ref{freqdef}) corresponding to the modes $\eta_i$. As a result, instead of our original $\hat{\cal H}_{int}$, it is sufficient to use
\beq\label{Hres}
\hat{\cal H}_{int}=\hspace{-6mm}\sum_{\om_1+\om_2=\om_3+\om_4} \hspace{-6mm}C_{\eta_1\eta_2\eta_3\eta_4} \ald_{\eta_1}  \ald_{\eta_2}  \al_{\eta_3}  \al_{\eta_4}.
\eeq
Within each $(N,E')$-level, this Hamiltonian is a finite-dimensional matrix that can be explicitly diagonalized. Implementing such diagonalization in practice has been discussed  in \cite{quantres} for somewhat simpler closely related systems (where the resonant structure of the sums is identical to the above, but the modes are labelled by a single nonnegative integer). Once the eigenvalues ${\cal E}^{(N,E')}_J$ within such finite-dimensional $(N,E')$-block have been found, at linear order in $\lambda$, the energy levels emanating from a given degenerate unperturbed $(N,E')$-level are simply
\beq
E_{(N,E'),J}=E'+\de N +\lambda \, {\cal E}^{(N,E')}_J.
\label{Eshift}
\eeq
Here, $J$ labels the different levels within the fine structure emanating from a given degenerate energy level of the non-interacting system specified by $(N,E')$.

We note that the Hamiltonian (\ref{Hres}) could be studied in its own right as a quartic dynamical system. Classical analogs of this {\it quantum resonant system}, with various different choices of the {\it interaction coefficients} $C$ (including the concrete problem we are considering here as a special case), have often surfaced in recent literature: resonant systems of gravitational AdS perturbations \cite{FPU,CEV, BMR,islands,returns,nrel} formulated in relation to the conjectured AdS instability \cite{BR,review}, resonant systems of nonlinear wave equations in AdS \cite{BKS,CF,BHP,BEL} and the related Gross-Pitaevskii equation for Bose-Einstein condensates \cite{GHT,BBCE,GGT,BBCE2}, a solvable model of turbulence in the form of a specific resonant system called the cubic Szeg\H o equation \cite{GG}, as well as studies of a large class of partially solvable resonant systems \cite{AO}. Quantum resonant systems, on the other hand, have been introduced and studied in \cite{quantres} from a perspective geared toward quantum chaos theory.

The energy shifts ${\cal E}^{(N,E')}_J$, which are effectively computable by finite matrix diagonalization, encode the physics of the problem. Our concrete and simple choice of the action (\ref{Lagr}) results in specific values of the interaction coefficients $C$ in (\ref{Hintdef}), which are  converted to specific energy shifts ${\cal E}^{(N,E')}_J$ by the above diagonalization procedure. Had we chosen a different interaction term in (\ref{Lagr}), the expressions for $C$, and hence the energy shifts, would have been different. One thus can effectively discriminate different theories in this approach, and to measure couplings. Including extra fields, which we shall briefly comment on below, would introduce extra oscillator variables in addition to $\al_\eta$. This would, however, simply result in an enlargement of the resonant Hamiltonian (\ref{Hres}) retaining its key features.

If one prepares a superposition of different energy eigenstates (\ref{Eshift}), different terms will oscillate through the usual quantum-mechanical phase factors $\exp[-iE_{(N,E'),J}\,t]$. Importantly, for $t\sim 1/\lambda$, the contribution of the interactions, which is  $\exp[-i\lambda \, {\cal E}^{(N,E')}_J t]$ becomes  of order 1, no matter how small $\lambda$ is. This is a mathematical expression of the resonant enhancement of small interactions in the AdS cavity. Note that is is wise to prepare superpositions of states emanating from the same $(N,E')$ energy level of the unperturbed system. In this way, rapid phase oscillations due to   $\exp[-i (E'+\de N) t]$ are identical for all terms of the superposition and will therefore not cause observable interference phenomena. Interference will emerge only at late times  $t\sim 1/\lambda$, and will specifically encode the information about the small interactions. Note that while energy shifts of order $\lambda$ exist in any cavity, AdS with its perfect resonant properties provides the luxury of highly degenerate unperturbed energy levels whose fine structure splitting is a powerful indicator of the properties of small interactions. (The spectrum of non-interacting fields in a generic cavity will by contrast be nondegenerate).

\vspace{3mm} 


\noindent{\bf Maximally rotating sectors:} In classical resonant systems corresponding to the Hamiltonian (\ref{Hres}), it is known that one can perform many interesting consistent truncations setting most of the mode amplitudes to zero \cite{BEL, BBCE,BBCE2}. For instance, one could keep only spherically symmetric modes. Such truncations do not in general work in the quantum version (\ref{Hres}) because of the uncertainty principle: in a quantum theory, nothing can be exactly zero, and quantum fluctuations of modes in their ground states will affect the other modes. There is, however, a particular `maximally rotating' truncation (whose analog is referred to as the Lowest Landau Level in the literature on Bose-Einstein condensates) which has a non-trivial analog in the corresponding quantum theory. We shall now display this truncation explicitly, as it is useful to keep in mind when contemplating the phenomenology of interacting particles in AdS.

At each frequency level (\ref{freqdef}), there is only one state with the maximal possible angular momentum $k$. To get this state, one sets $l=k$ (because multiplets with $l<k$ do not contain states with angular momentum $k$) and $n=0$. For this state $\om -\de=k$. Because of the rotational symmetry, the Hamiltonian (\ref{Hres}) conserves the angular momentum
\beq
\hat K=\sum_{nlk} k\,\ald_{nlk}\al_{nlk},
\eeq
in addition to (\ref{Edef})-(\ref{Epdef}).
Of course, the Hamiltonian is block-diagonal in $K$ and can only have nonvanishing matrix elements between states with the same $K$. Imagine, we choose to consider states with $K=E'$. Such states can only have nonvanishing occupation numbers $n_\eta$ for modes with  $\om -\de=k$, i.e., precisely the maximally rotating states we have described above. Any individual particle occupying a different AdS mode will automatically ensure that $K<E'$ for the entire multi-particle state, so such states are not in our $K=E'$ block, and dynamically disconnected from it with respect to (\ref{Hres}). One can then simply forget about all the remaining modes, and label the maximally rotating $\alpha$'s by a single nonnegative integer $k$. The resulting system, `truncated' to the maximally rotating sector, is
\beq\label{HresMR}
\hat{\cal H}^{(MR)}_{int}=\hspace{-6mm}\sum_{k_1+k_2=k_3+k_4} \hspace{-6mm}C_{k_1 k_2 k_3 k_4} \ald_{k_1}  \ald_{k_2}  \al_{k_3}  \al_{k_4}.
\eeq
It is known from \cite{BEL} that the classical analog of this maximally rotating truncation in AdS has many special properties (and admits some explicit analytic solutions despite the presence of strong nonlinearites). It also belongs to a large class of `partially solvable' resonant systems discussed in \cite{AO} and displaying extra conserved quantities. On the other hand, the quantum resonant system (\ref{HresMR}) is of a general form studied in detail in \cite{quantres}. These structures allow analyzing the levels of (\ref{HresMR}) in great detail. (Similar considerations for the related resonant system of the two-dimensional Gross-Pitaevskii equation and its Lowest Landau Level truncation shall be presented at length in a forthcoming publication \cite{shifts}.) Having an explicitly tractable maximally rotating sector for weakly interacting particles in AdS is certainly a welcome feature.

\vspace{3mm} 

\noindent{\bf Multiple particle species:} We comment a bit further on what happens when there are many particle species involved, a common situation in particle physics. Each particle species will experience resonant enhancement of self-interactions over long times in the manner described above. We emphasize once again that the presence of resonances is universal for all particle masses and spins and is deeply rooted in the structure of the symmetry group of AdS -- see, e.g., \cite{EN}.

What about the interactions of different particle species? They are also enhanced, but for generic masses, this only happens for processes that do not change particle numbers. This is because different masses will correspond to different, and  in general not commensurate, values of the real parameter $\de$ in (\ref{freqdef}). So the frequencies of species 1 will be in the form $\de_1 +\mbox{integer}$ and for the species 2, $\de_2+\mbox{integer}$. Since the sum of frequencies before and after the reaction must be the same, for non-commensurate $\de_1$ and $\de_2$, we must have the number of particles of species 1 and 2 conserved independently.

If one can control the AdS radius $L$, one could tune $\de_1$ and $\de_2$ to be commensurate as $m^2$ gets replaces by $m^2L^2$ in (\ref{freqdef}) -- or, better still, tune them into integer values. In this case, resonant channels are open for particle production. One would still have to make sure, however, that selection rules analogous to \cite{Yang, EN} do not close the operation of these open resonant channels. This depends on the types of particles involved, and on the structure of the interaction terms. Whether or not one can control $L$, on the other hand, depends on the concrete implementation of the AdS cavity, which we are not discussing here. Overall, the idea of resonantly enhancing small interaction in the AdS setup seems to be more suited for accurate measurements of very small couplings between known particle species than for discovery of new particles (as resonant production of particles is difficult to arrange). One may nonetheless continue looking for further ingenious modifications of the elementary setup we have presented here that would be beneficial for particle production studies. 

\vspace{3mm} 


\noindent{\bf Discussion:} We have laid out a hypothetical scenario where a particle experimentalist would have access to Anti-de Sitter spacetime, and the advantages of studying small couplings in this setting. Being a perfectly resonant cavity for fields of all masses and spins, AdS dramatically enhances small interactions if on waits long enough. The resonances between the AdS normal modes manifest themselves as extremely high degeneracies of the energy levels in the absence of interactions. Interactions split these levels, and one could extract information about the underlying couplings by studying the spectroscopy of the fine structure emanating from each unperturbed level. Over long times, the energy shifts induce large phase drifts of the different wavefunction components, enhancing arbitrarily small interactions to effects of order 1.

Energy shifts due to interactions are, of course, not unique to AdS and exist for any weakly interacting fields in a cavity. In a generic cavity, however, unperturbed energy levels are non-degenerate and acquire small shifts individually when the interactions are turned on. These tiny shifts would have to be disentangled in measurements from the large and irregular spacings between the unperturbed energy levels. By contrast, the special symmetry structure of AdS mandates highly degenerate evenly spaced energy levels for non-interacting fields. The level splitting due to interactions is controlled by neat algebraic structures in the spirit of \cite{quantres, shifts}. If one prepares a mixture of energy levels emanating from a single unperturbed level, the fast free-field evolution becomes an irrelevant common phase, while the relative phases of the fine structure components evolve very slowly and directly encode the information about the interactions. None of these features would be available in a non-resonant cavity, while a less resonant cavity will correspondingly supply fewer advantages in terms of how the interactions are encoded in the multi-particle spectroscopy.

Resonant spacetimes other than AdS do exist. For instance, one can construct large classes of spacetimes which are resonant with respect to scalar fields of particular masses \cite{resspa}. There are also situations where a highly resonant spectrum emerges specifically for spherically symmetric normal modes, but not for other modes \cite{cavity,R2}. It seems likely that AdS (being extremely special from the symmetry standpoint) is the only spacetime resonant for all masses and spins, though we are not aware of a proof. 

Our Universe is not an AdS spacetime for all we know. (In fact, on large cosmological scales, de Sitter spacetime seems a much better approximation, and its phenomenology is very different from what we have been discussing here.) Could one create AdS in a lab? If one manages to lower the vacuum energy density in a finite region of space to a constant time- and position-independent value, approximately AdS spacetime will emerge there. In order to reproduce the global properties of AdS well (importantly, its resonant structure) the size of the space-time region must be much bigger than the AdS radius. If this condition is met, the arrangements at the boundary of our spatial region are not particularly important since the AdS-like gravitational field we have created will hold the particles in the interior. Massless particles can still travel far from the interior, and one could set up a reflective boundary for them enclosing our AdS-like experimental domain (which mimics the reflective boundary conditions at infinity for massless fields in AdS proper). If created, such a cavity would approximate with a large precision the phenomenology outlined in our treatment. Being able to control the AdS radius (set by the value of the negative vacuum energy) would provide extra advantages in analyzing particle interactions. Admittedly, we do not have a way at this moment to manipulate the vacuum energy to achieve such experimental goals, but who knows what the future will bring. We shall refrain from speculating further...\vspace{3mm}


\noindent{\bf Acknowledgments:} The idea of this essay emerged while giving a mathematical physics talk based on \cite{quantres} for an audience dominated by particle experimentalists at the HEPMAD18 conference in Antananarivo, Madagascar. I thank the organizers and the participants for putting together this unusual meeting in a remarkable place. Research presented here has been funded by the CUniverse research promotion project (CUAASC) at Chulalongkorn University.

\end{document}